# The Sine Gordon Model: Perturbation Theory and Cluster Monte Carlo

M. Hasenbusch[a], M. Marcu[b], and K. Pinn[c]

[a] *Theory Division, CERN*
*CH–1211 Geneva 23, Switzerland* [1]

after September 30, 1994:
DAMTP, Silver Street,
Cambridge, CB3 9EW, England

[b] *Racah Institute of Physics*
*Hebrew University, 91904 Jerusalem, Israel* [2]

[c] *Institut für Theoretische Physik I, Universität Münster*
*Wilhelm-Klemm-Str. 9, D-48149 Münster, Germany* [3]

**Abstract**

We study the expansion of the surface thickness in the 2-dimensional lattice Sine Gordon model in powers of the fugacity $z$. Using the expansion to order $z^2$, we derive lines of constant physics in the rough phase. We describe and test a VMR cluster algorithm for the Monte Carlo simulation of the model. The algorithm shows nearly no critical slowing down. We apply the algorithm in a comparison of our perturbative results with Monte Carlo data.

[1] e–mail: hasenbus @surya11.cern.ch
[2] e-mail: marcu @decscc.tau.ac.il
[3] e-mail: pinn @yukawa.uni-muenster.de

# 1 Introduction

The 2-dimensional lattice Sine Gordon (SG) model is a prominent member of a large class of solid-on-solid (SOS) models that are believed to undergo a roughening transition of the Kosterlitz-Thouless type [1, 2]. The model can be considered as a Gaussian model with a perturbation that is periodic in the real variables $\varphi_x$. The Hamiltonian is

$$H(\varphi) = \frac{1}{2\beta} \sum_{<x,y>} (\varphi_x - \varphi_y)^2 - V(\varphi). \tag{1}$$

The sum runs over all nearest neighbour pairs $<x, y>$ in the 2-dimensional lattice, and

$$V(\varphi) = z \sum_x \cos(2\pi\varphi_x). \tag{2}$$

We have chosen units such that the Boltzmann factor is $\exp(-H)$.

Compared to other SOS models[4] the SG model is very suitable for analytical calculations, e.g., expansions in the fugacity $z$.[5]

The variables $\varphi_x$ have a natural interpretation as height variables. For sufficiently small temperature $\beta$ a typical $\varphi$-configuration describes a more or less smooth surface. In the thermodynamic limit the surface gets localized and has a finite thickness. The model is in the *smooth* phase.

For large $\beta$, however, the surface can freely wander. Furthermore, the surface thickness squared diverges logarithmically when the area becomes large. The model is in the *rough* phase.

The two phases are separated by a phase transition of infinite order, the roughening transition. The transition occurs at a critical line $\beta_c(z)$.

To make things more precise, we define the surface thickness $\sigma$. On a finite periodic square lattice $\Lambda$ with $N = L \times L$ sites, let

$$\sigma^2 = \frac{1}{N^2} \sum_{x,y} \langle (\varphi_x - \varphi_y)^2 \rangle = 2\langle (\varphi_x - \Phi)^2 \rangle. \tag{3}$$

$\Phi$ is the average of $\varphi$ over the entire lattice,

$$\Phi = \frac{1}{N} \sum_{x \in \Lambda} \varphi_x. \tag{4}$$

A proof is presented in appendix 1 that for $z = 0$ and $L \to \infty$,

$$\sigma^2 \to \frac{\beta}{\pi} \ln L + c, \tag{5}$$

---

[4]For a Monte Carlo renormalization group investigation of several SOS models see ref. [3].

[5]The name fugacity for the coupling constant $z$ will become clear later when we consider the Coulomb gas representation of the Sine Gordon model, see also ref. [4].

where $c$ is some constant.

The surface thickness can be used to distinguish between the two phases of the Sine Gordon model: In the limit $L \to \infty$, for nonzero $z$,

$$\sigma^2 \to \begin{cases} \frac{\beta_{\text{eff}}(z)}{\pi} \ln L + c', & \text{if } \beta \geq \beta_c(z), \\ \text{finite}, & \text{else}. \end{cases} \qquad (6)$$

This equation says that in the rough (large $\beta$) phase, the long distance behaviour of the theory is that of a Gaussian model ($z = 0$) with infinite correlation length, but with a changed temperature $\beta_{\text{eff}}$ that depends on the fugacity $z$.

It is known that for $z \to 0$, the critical temperature goes to $2/\pi$, see, e.g., [4]. Accordingly, the *critical line* is given as the set of points $(z, \beta)$ with $\beta_{\text{eff}}(z) = 2/\pi$. For every $z$, the critical point $\beta_c(z)$ is the smallest value of $\beta$ such that the surface thickness diverges for $L \to \infty$. At $\beta = \beta_c(z)$, the asymptotic ratio $\sigma^2/\ln(L)$ jumps from $2/\pi^2$ to zero. More generally, we shall define *lines of constant physics* in the rough phase of the model as those lines in the $(z, \beta)$-plane that belong to the same value of $\beta_{\text{eff}}$. Models that lie on the same line have identical long distance behaviour (which is Gaussian in this particular model).[6]

In this paper, we shall derive an approximation of the lines of constant physics from the perturbative expansion of the surface thickness $\sigma$ to second order in $z$.

It is always interesting to compare analytical results with numerical ones, especially when approximations are made in the analytical calculations. To make this comparison we employed a cluster algorithm for the simulation of the SG model. This algorithm is a new member in a class of cluster algorithms that we call VMR (valleys-to-mountains-reflection) algorithms [5]. In this paper, we shall describe the algorithm and study its dynamical critical properties. We use it to produce estimates for the surface thickness and other quantities that we can compare with the perturbative results.

This paper is organized as follows. In section 2 we develop the perturbation theory for the surface thickness. Section 3 is devoted to the derivation of a formula for the lines of constant physics in the rough phase. We then turn in section 4 to the description and study of the VMR cluster algorithm for the SG model. In section 5 we compare the analytical results with those obtained with the Monte Carlo algorithm.

## 2 Perturbation Theory for the Surface Thickness

Expectation values (correlation functions) of observables in the Sine Gordon model are defined through

$$\langle \mathcal{O}(\varphi) \rangle = \frac{\int \prod_x d\varphi_x \exp\left[-H(\varphi)\right] \mathcal{O}(\varphi)}{\int \prod_x d\varphi_x \exp\left[-H(\varphi)\right]}. \qquad (7)$$

---

[6]According to D. Stauffer, one might call these lines also "isotrachs", a term composed from greek words, which means "lines of constant roughness".

The Hamiltonian $H(\varphi)$ is invariant under shifts $\varphi_x \to \varphi_x + n$ for all $x \in \Lambda$, where $n$ is an integer constant. As a consequence, expectation values are defined only for observables $\mathcal{O}(\varphi)$ that have the same global symmetry: In this case the infinite contribution from the zero mode associated with the symmetry is exactly cancelled in eq. (7).[7]

For the perturbation theory to be done in this section, it is convenient to rewrite eq. (7) with the help of a Gaussian measure. An integration measure $d\mu_{\beta C}(\varphi)$ is defined through its "generating functional"

$$\int d\mu_{\beta C}(\varphi) \exp[i(k,\varphi)] = \begin{cases} \exp\left[-\tfrac{1}{2}\beta(k,Ck)\right] & \text{, if } \sum_x k_x = 0 \\ 0 & \text{, else .} \end{cases} \tag{8}$$

The covariance $C$ is a matrix with elements

$$\begin{aligned} C_{xy} &= \frac{1}{N} \sum_{p \neq 0} \frac{e^{ip(x-y)} - 1}{\hat{p}^2}, \\ \hat{p}^2 &= 4 - 2\cos p_1 - 2\cos p_2, \end{aligned} \tag{9}$$

where the $p_i$, $i = 1, 2$, are summed over the values $\{0, ..., L-1\} \cdot (2\pi/L)$. The scalar product is defined by

$$(\phi, \psi) = \sum_x \phi_x \psi_x . \tag{10}$$

It is shown in appendix 2 that the expectation value eq. (7) can be rewritten as

$$\langle \mathcal{O}(\varphi) \rangle = \frac{\int d\mu_{\beta C}(\varphi) \exp[V(\varphi)] \mathcal{O}(\varphi)}{\int d\mu_{\beta C}(\varphi) \exp[V(\varphi)]} . \tag{11}$$

The calculation of the interface thickness is started with the observation that for arbitrary observables $\mathcal{O}(\varphi)$,

$$\begin{aligned} \langle \mathcal{O}(\varphi) \rangle &= \frac{\int d\mu_{\beta C}(\varphi) \exp[V(\varphi)] \mathcal{O}(\varphi)}{\int d\mu_{\beta C}(\varphi) \exp[V(\varphi)]} \\ &= \frac{\int d\Phi \int d\mu_{\beta C}(\varphi) \delta(\Phi - \frac{1}{N}\sum_x \varphi_x) \exp[V(\varphi)] \mathcal{O}(\varphi)}{\int d\Phi \int d\mu_{\beta C}(\varphi) \delta(\Phi - \frac{1}{N}\sum_x \varphi_x) \exp[V(\varphi)]} \\ &= \frac{\int d\Phi \int d\mu_{\beta \Gamma}(\varphi) \exp[V(\Phi + \varphi)] \mathcal{O}(\Phi + \varphi)}{\int d\Phi \int d\mu_{\beta \Gamma}(\varphi) \exp[V(\Phi + \varphi)]} . \end{aligned} \tag{12}$$

Here, the Gaussian measure with covariance $\Gamma$ is defined through

$$\int d\mu_\Gamma(\varphi) \mathcal{O}(\varphi) = \lim_{M \to 0} \frac{\int d\mu_{v_M}(\varphi) \delta(\sum_x \varphi_x) \mathcal{O}(\varphi)}{\int d\mu_{v_M}(\varphi) \delta(\sum_x \varphi_x)} , \tag{13}$$

---
[7]The **Z**-symmetry of the model is broken in the smooth phase, $\beta < \beta_c(z)$. Here, in the thermodynamic limit, expectation values of observables that are not **Z**-invariant, are defined.

where[8] $v_M = (-\Delta + M^2)^{-1}$. It is not difficult to demonstrate that

$$\Gamma_{xy} = \int d\mu_\Gamma(\varphi)\, \varphi_x \varphi_y = \frac{1}{N} \sum_{p \neq 0} \frac{e^{ip(x-y)}}{\hat{p}^2}\,. \tag{14}$$

Correlation functions as occuring in eq. (12) can be computed with the help of the generating functional

$$\int d\mu_{\beta\Gamma}(\varphi) \exp[i(k,\varphi)] = \exp\left[-\tfrac{1}{2}\beta(k,\Gamma k)\right]\,. \tag{15}$$

The "effective observable" for the surface thickness is independent of $\Phi$:

$$\mathcal{O}(\Phi + \varphi) = 2\,\varphi_x^2\,. \tag{16}$$

We therefore have

$$\sigma^2 = 2\, \frac{\int d\mu_{\beta\Gamma}(\varphi)\, \varphi_x^2 \int d\Phi\, \exp[V(\Phi + \varphi)]}{\int d\mu_{\beta\Gamma}(\varphi) \int d\Phi\, \exp[V(\Phi + \varphi)]}\,. \tag{17}$$

We expand the Boltzmann factor:

$$\exp[V(\Phi + \varphi)] = \sum_{n \geq 0} \frac{z^n}{n!} \sum_{x_1} \ldots \sum_{x_n} \cos 2\pi(\varphi_{x_1} + \Phi) \ldots \cos 2\pi(\varphi_{x_n} + \Phi) \tag{18}$$

$$= \sum_{n \geq 0} \left(\frac{z}{2}\right)^n \frac{1}{n!} \sum_{x_1, s_1} \ldots \sum_{x_n, s_n} \exp\left[i 2\pi \left(\sum_{i=1}^n s_i \cdot (\varphi_{x_i} + \Phi)\right)\right]\,. \tag{19}$$

The "charges" $s_i$ are summed over the values $\pm 1$. The $\Phi$-integral leads to a $\delta$-function that forces the sum of the $s_i$ to be zero (neutrality condition). We shall indicate this constraint with a prime at the sums in the equations to follow. The Gaussian integrations in eq. (17) can now be done, leading to the representation

$$\sigma^2 = 2\beta\Gamma_{oo} - K\, \frac{\sum_{n \geq 0} \left(\frac{z}{2}\right)^n Z'_n}{\sum_{n \geq 0} \left(\frac{z}{2}\right)^n Z_n}\,. \tag{20}$$

Here, we have defined

$$Z'_n = \frac{1}{n!} {\sum_{x_1, s_1}}' \ldots {\sum_{x_n, s_n}}' \exp\left(-\tilde{\beta} \sum_{i<j}^n s_i C_{x_i x_j} s_j\right) \left(\sum_{i=1}^n C_{o x_i} s_i\right)^2\,, \tag{21}$$

and

$$Z_n = \frac{1}{n!} {\sum_{x_1, s_1}}' \ldots {\sum_{x_n, s_n}}' \exp\left(-\tilde{\beta} \sum_{i<j}^n s_i C_{x_i x_j} s_j\right)\,. \tag{22}$$

---
[8]For the definition of the lattice Laplacian see appendix 2.

Furthermore, we have introduced the abbreviations

$$\tilde{\beta} = (2\pi)^2 \beta$$
$$K = \frac{2\tilde{\beta}^2}{(2\pi)^2}. \tag{23}$$

In the derivation of eqs. (21) and (22) we exploited the fact that for neutral charge configurations $s$,

$$\sum_{i,j}^{n} s_i \Gamma_{x_i x_j} s_j = 2 \sum_{i<j}^{n} s_i C_{x_i x_j} s_j, \tag{24}$$

and

$$\sum_{i}^{n} \Gamma_{o x_i} s_i = \sum_{i}^{n} C_{o x_i} s_i. \tag{25}$$

Eq. (20) is the Coulomb gas representation of the surface thickness. The gas is a neutral gas with integer charges $\pm 1$. A lattice site can be occupied by more than one charge. The representation is in the grand canonical ensemble, and $z$ obviously plays the role of the fugacity for the charges $s_i$. A small $z$ means that the dominant contributions to the sum come from systems with only a few charges present (dilute Coulomb gas). Note that it follows from this representation that a finite fugacity always makes the surface thickness smaller, because only positive terms are subtracted from the $z = 0$ result.[9] Furthermore, together with eq. (6) it follows that $\beta_{\text{eff}}$ is smaller than $\beta$.

We now determine the coefficients in the expansion

$$\sigma^2 = \sum_{n \geq 0} \left(\frac{z}{2}\right)^n \sigma_n^2. \tag{26}$$

One finds

$$\begin{aligned}
\sigma_n^2 &= 0 \quad \text{for } n \text{ odd} \\
\sigma_0^2 &= 2\beta \Gamma_{oo} \\
\sigma_2^2 &= -K Z_2' \\
\sigma_4^2 &= -K(Z_4' - Z_2' Z_2) \\
\sigma_6^2 &= -K(Z_6' - Z_2' Z_4 - Z_2' Z_4 + Z_2' Z_2^2) \\
\sigma_8^2 &= \ldots
\end{aligned} \tag{27}$$

Let us explicitly write down $Z_2'$, $Z_2$, and $Z_4'$:

$$\begin{aligned}
Z_2' &= \sum_{x,y} \exp\left(\tilde{\beta} C_{xy}\right) (C_{ox} - C_{oy})^2 \\
&= \sum_{x \neq 0} \exp\left(\tilde{\beta} C_{ox}\right) \sum_{y} (C_{o,x+y} - C_{o,y})^2, \tag{28}
\end{aligned}$$

---

[9]We here assume that the infinite sum over $n$ converges. This is probably the case in the rough phase.

$$Z_2 = L^2 \sum_x \exp\left(\tilde{\beta} C_{ox}\right) . \tag{29}$$

Finally,

$$\begin{aligned}
Z'_4 &= \frac{1}{4} \sum_{x_1, x_2, x_3, x_4} \exp\left[-\tilde{\beta}\left(C_{x_1 x_2} - C_{x_1 x_3} - C_{x_1 x_4} - C_{x_2 x_3} - C_{x_2 x_4} + C_{x_3 x_4}\right)\right] \cdot \\
&\quad \cdot (C_{ox_1} + C_{ox_2} - C_{ox_3} - C_{ox_4})^2
\end{aligned} \tag{30}$$

Of course, using translational invariance of $C$, the expression for $Z'_4$ can be transformed in a similar way as for $n = 2$.

## 3  Lines of Constant Physics

In the rough phase the behaviour of the 2-dimensional SG model at large distance is that of a Gaussian model ($z = 0$), but with an effective $\beta_{\text{eff}} < \beta$. For the surface thickness this means that in the limit $L \to \infty$,

$$\sigma^2 \to \frac{\beta_{\text{eff}}(z)}{\pi} \ln L + c' . \tag{31}$$

We define lines of constant physics as the sets of points $(z, \beta)$ that belong to the same $\beta_{\text{eff}}$. Constant physics here means that the models lying on the same line have identical long distance behaviour. In this section we will use the expansion of the surface thickness to second order in $z$ to determine the lines of constant physics. With eq. (28) and the observation that

$$\sum_y (C_{o,x+y} - C_{o,y})^2 = -2 \frac{1}{N} \sum_{p \neq 0} \frac{e^{ipx} - 1}{(\hat{p}^2)^2} \tag{32}$$

one arrives at

$$\sigma^2 = \sigma_o^2 + z^2 \frac{\tilde{\beta}^2}{(2\pi)^2} \sum_{x \neq 0} \exp\left(\tilde{\beta} C_{ox}\right) \frac{1}{N} \sum_{p \neq 0} \frac{e^{ipx} - 1}{(\hat{p}^2)^2} + O(z^4) . \tag{33}$$

The first observation is that the $O(z^2)$ term of the $z = 0$ surface thickness also behaves like $\ln L$ for large $L$. To show this, we interchange the order of summation,

$$\sum_{x \neq 0} \exp\left(\tilde{\beta} C_{ox}\right) \frac{1}{N} \sum_{p \neq 0} \frac{e^{ipx} - 1}{(\hat{p}^2)^2} = \frac{1}{N} \sum_{p \neq 0} \frac{1}{\hat{p}^2} \underbrace{\sum_{x \neq 0} \exp\left(\tilde{\beta} C_{ox}\right) \frac{e^{ipx} - 1}{\hat{p}^2}}_{D_p} . \tag{34}$$

Note that $D_p = A + O(p^2)$, and $A$ has an infinite volume limit:

$$A_\infty = \lim_{L\to\infty}\lim_{p\to 0} D_p = -\tfrac{1}{2} \lim_{p\to 0} \sum_{x\neq 0} \exp\left(\tilde{\beta} C_{ox}\right) \frac{(px)^2}{\hat{p}^2}$$
$$= -\tfrac{1}{4} \sum_{x\neq 0} \exp\left(\tilde{\beta} C_{ox}\right) x^2 . \qquad (35)$$

The $x$-sum now runs over the infinite lattice. The $O(p^2)$-terms in $D_p$ will lead to corrections of $\sigma_o^2$ that remain finite in the infinite volume limit. The limit $p \to 0$ in eq. (35) is a bit subtle. Note that the limit of $(px)^2/\hat{p}^2$ alone strongly depends on the direction in that the limit is taken in the 2-dimensional $p$-space. However, the limit is to be performed under the sum over $x$. Now divide this sum into a sum over equivalence classes $X$ of points $x$ on that $C_{ox}$ takes the same value (these points are actually connected by lattice symmetries). Sum $(px)^2/\hat{p}^2$ over $x \in X$ and then let $p \to 0$. Then the limit becomes independent of the direction in that the limit is taken, and one arrives at the result given in eq. (35).

The large $L$ approximation is thus

$$\sigma^2 \to \sigma_o^2 + z^2 \frac{\tilde{\beta}^2}{(2\pi)^2} \underbrace{\frac{1}{N} \sum_{p\neq 0} \frac{1}{\hat{p}^2}}_{\Gamma_{oo}} A_\infty + \text{ const } + O(z^4). \qquad (36)$$

We apply the approximation formula [8]

$$C_{ox} \simeq -\frac{1}{2\pi}\left(\ln|x| + \tfrac{3}{2}\ln 2 + \gamma\right), \qquad (37)$$

where $\gamma = 0.5772...$ denotes Euler's constant. The approximation for $A_\infty$ then is

$$A_\infty \approx -\tfrac{1}{4} \exp\left[-2\pi\beta(\tfrac{3}{2}\ln 2 + \gamma)\right] \underbrace{\sum_{x\neq 0} |x|^{2-2\pi\beta}}_{E_2(\pi\beta-1)} . \qquad (38)$$

$E_2$ denotes the Epstein-$\zeta$-function [9].

Note that replacing the "true" Coulomb propagator by its approximation eq. (37) will change $A_\infty$ by a small finite amount (the approximation of the propagator has errors that decay like $1/|x|$). However, $A_\infty$ diverges for $\beta \to 2/\pi$. Since the divergence comes from the large $x$ contributions, the finite errors from the small distances can be neglected in this limit which will be of importance later.
Recall that $\sigma_o^2 = 2\beta \Gamma_{oo}$. Therefore

$$\sigma^2 \to 2\Gamma_{oo}\left(\beta - z^2 \tfrac{1}{8}(2\pi)^2 \beta^2 \exp\left[-2\pi\beta(\tfrac{3}{2}\ln 2 + \gamma)\right] E_2(\pi\beta-1)\right) + \text{ const } + O(z^4). \qquad (39)$$

If we define a $\Delta\beta$ function through

$$\beta_{\text{eff}} = \beta - z^2 \Delta\beta(\beta) + O(z^4), \qquad (40)$$

we find
$$\Delta\beta(\beta) = \frac{1}{8}(2\pi)^2\beta^2 \exp\left[-2\pi\beta(\tfrac{3}{2}\ln 2 + \gamma)\right] E_2(\pi\beta - 1). \tag{41}$$

Let us now introduce a new variable
$$x = \beta - \frac{2}{\pi}. \tag{42}$$

$x$ measures the distance from the critical point at vanishing fugacity $z$. For $\Delta x(x)$ one obtains
$$\Delta x(x) = \frac{1}{8}(2\pi)^2 \left(\frac{2}{\pi} + x\right)^2 \exp\left(-2\pi\left(\frac{2}{\pi} + x\right)\left(\tfrac{3}{2}\ln 2 + \gamma\right)\right) E_2(1 + \pi x). \tag{43}$$

We are now able to obtain the lines of constant physics. They will be parameterized in the form $z(x, x_{\text{eff}})$, where $x_{\text{eff}} = \beta_{\text{eff}} - \frac{2}{\pi}$:
$$z(x, x_{\text{eff}}) = \left(\frac{x - x_{\text{eff}}}{\Delta x(x)}\right)^{1/2}. \tag{44}$$

In figure 1, we show the lines of constant physics for $x_{\text{eff}} = 0.0\ldots0.09$. The case $x_{\text{eff}} = 0.0$ corresponds to the critical line. For the numerical evaluation of $\Delta x(x)$ we used that [9] the Epstein $\zeta$-function $E_2$,
$$E_2(s) = \sum_{x \neq 0} \frac{1}{(x_1^2 + x_2^2)^s}, \tag{45}$$

can be written as the product
$$E_2(s) = 4\zeta_R(s)\beta(s), \tag{46}$$

where $\zeta_R(s)$ denotes the Riemann $\zeta$-function,
$$\zeta_R(s) = \sum_{n=1}^{\infty} n^{-s}, \tag{47}$$

and [10]
$$\beta(s) = \sum_{n=0}^{\infty} (-1)^n (2n+1)^{-s} = \frac{1}{2\Gamma(s)} \int_0^{\infty} dt \, \frac{t^{s-1}}{\cosh(t)}. \tag{48}$$

Figure 1 looks like the famous flow diagram of Kosterlitz and Thouless. The strong similarity becomes more evident if we expand $z(x, x_{\text{eff}})$ around $x = 0$. Using the expansions
$$\zeta_R(s) = \frac{1}{s-1} + \gamma + O(s-1), \tag{49}$$

and [11]
$$\beta(s) = \frac{\pi}{4}\left(1 + \left(\gamma + \ln 2\pi + 2\ln\frac{\Gamma(3/4)}{\Gamma(1/4)}\right)(s-1)\right) + O\left((s-1)^2\right), \tag{50}$$

we find
$$E_2(1 + \pi x) = \frac{1}{x} + \kappa + O(x), \qquad (51)$$
with
$$\kappa = \pi \left(2\gamma + \ln 2\pi + 2\ln \frac{\Gamma(3/4)}{\Gamma(1/4)}\right) = 2.5849817.... \qquad (52)$$
If we plug in all our material, we find
$$z(x, x_{\text{eff}}) = S\sqrt{x(x - x_{\text{eff}})}\left(1 + O(x)\right), \qquad (53)$$
where $S = \exp((5/2)\ln 2 + 2\gamma) \approx 17.945$. The critical line in the vicinity of the fixed point $(z = 0, x = 0)$ is thus
$$z(x, 0) = S x \left(1 + O(x)\right). \qquad (54)$$
For a comparison: The solutions of the Kosterlitz-Thouless equations in the rough phase are of the form
$$Y = \sqrt{E + X^2}, \qquad (55)$$
with $E < 0$. $Y$ and $X$ are proportional to the fugacity and to $\beta - 2/\pi$, respectively.[10] The case $E = 0$ corresponds to the critical line.

## 4  Cluster Algorithm for the Sine Gordon Model

### 4.1  VMR for the Sine Gordon Model

Cluster algorithms have proven successful in fighting critical slowing down (CSD) in the simulation of spin models. Cluster algorithms first occured in the pioneering work of Swendsen and Wang [6] in the context of the Ising model.

Let us first describe the Swendsen-Wang algorithm as applied to an Ising model with partition function
$$Z = \sum_{\sigma_x = \pm 1} \exp\left(\sum_{<x,y>} K_{<x,y>} \sigma_x \sigma_y\right), \qquad (56)$$
where $<x, y>$ is the bond connecting the sites $x$ and $y$.

In the first step of the algorithm all bonds of the lattice are deleted or frozen. A bond $<x, y>$ is deleted with probability
$$p_{d<x,y>} = \begin{cases} \exp(-K_{<x,y>}(1 + \sigma_x\sigma_y)) & \text{if} \quad K_{<x,y>} > 0, \\ \exp(+K_{<x,y>}(1 - \sigma_x\sigma_y)) & \text{if} \quad K_{<x,y>} \leq 0, \end{cases} \qquad (57)$$

---

[10]Note, however, that the corresponding constants are nonuniversal, i.e. depending on the cutoff scheme. They can not be determined by a trivial computation.

or frozen with probability $p_{f<x,y>} = 1 - p_{d<x,y>}$. In a second step, the clusters of sites connected by frozen bonds are identified. Finally, each cluster is flipped with probability one half.

Wolff proposed the so called single cluster version of this algorithm [7]. Here in one update step only one cluster is built. One randomly selects a seed $x_o$ for this cluster and connects neighbouring sites with the above given probability $p_f$ to this site. This procedure is continued with the neighbours of the new sites until no further sites join the cluster. Then one flips the spins in this cluster with probability one.

Cluster algorithms for the simulations of SOS models models were proposed in [5]. The idea for these algorithms came from the picture of SOS configurations as landscapes with hills and valleys. Large scale changes of a configuration are done by choosing a horizontal reflection plane, considering the connected regions above the plane (hills) and below the plane (valleys), and reflecting them through the plane independently, with an appropriate probability.

We shall describe the details of the algorithm in the language of the Discrete Gaussian (DG) model. The modifications necessary for the Sine Gordon model will be given later. Note that the DG model might be considered as the $z \to \infty$ limit of the SG model. The DG model has the partition function

$$Z = \sum_{\{h\}} \exp\left(-K^{DG} \sum_{<x,y>} (h_x - h_y)^2\right), \tag{58}$$

where $h_x$ is an integer, and the sum runs over nearest neighbour pairs.

Let us denote the height of the horizontal reflection plane by $M$. A reflection of $h_x$ with respect to $M$ means

$$h_x \to 2M - h_x. \tag{59}$$

Obviously, $M$ has to be either an integer or a half-integer. The reflection can be expressed in terms of embedded Ising variables, $s_x = \pm 1$, defined by the decomposition

$$h_x = s_x |h_x - M| + M. \tag{60}$$

We get the partition function of the embedded Ising model by inserting eq. (60) in eq. (58),

$$Z_{\text{eff}} = \text{const} \sum_{s_x = \pm 1} \exp\left(\sum_{<x,y>} K_{<x,y>} s_x s_y\right), \tag{61}$$

with

$$K_{<x,y>} = 2K^{DG} |h_x - M| |h_y - M|. \tag{62}$$

Hence we get from eq. (57)

$$p_{d<x,y>} = \exp\left(-2K^{DG} |h_x - M| |h_y - M| (1 - s_x s_y)\right). \tag{63}$$

It is important that the reflection plane lies with reasonable probability within the vertical bounds of the SOS surface. In [5], we introduced two different types of algorithms to ensure this.[11] In the *H-algorithm* one chooses $M$ in the neighbourhood of the height of the seed spin $h_{x_o}$. More precisely, $M$ is selected with a priori probability

$$P(M) = \begin{cases} \frac{1}{2} & \text{for} \quad M = h_{x_o} \pm \frac{1}{2}, \\ 0 & \text{else}. \end{cases} \qquad (64)$$

The H-algorithm alone is not successful in eliminating CSD. The situation is improved by including reflections with respect to an integer reflection plane $M$. In the *I-algorithm*, one selects an integer reflection plane $M$ by simply setting it equal to the height variable at a randomly chosen site not equal to the seed:

$$M = h_{y_o}, \quad y_o \neq x_o. \qquad (65)$$

Since the I-algorithm is not ergodic, we combined it with the H-algorithm. For the combined algorithm we did not find any CSD in the simulation of the DG model both in the rough and in the smooth phase.

We now turn to the SG model. The Hamiltonian is

$$H(\varphi) = \frac{1}{2\beta} \sum_{<x,y>} (\varphi_x - \varphi_y)^2 - z \sum_x \cos(2\pi\varphi_x). \qquad (66)$$

We use the embedding of Ising variables

$$\varphi_x = s_x |\varphi_x - M| + M. \qquad (67)$$

Because of its symmetry properties the SG potential remains unchanged if we perform cluster flips with respect to a reflection plane $M$ that is either integer or half integer. Our version of the H-algorithm for the SG model is specified by

$$P(M) = \begin{cases} \frac{1}{2} & \text{for} \quad M = \text{nint}(\varphi_{x_o}) \pm \frac{1}{2}, \\ 0 & \text{else}. \end{cases} \qquad (68)$$

and for the I-algorithm by

$$M = \text{nint}(\varphi_{y_o}), \quad y_o \neq x_o, \qquad (69)$$

where $y_o$ is again a randomly chosen site not equal to the seed $x_o$. Furthermore, $\text{nint}(\varphi_x)$ is the integer $n$ such that $|n - \varphi_x|$ is minimal. The delete probabilities are now given by

$$p_{d<x,y>} = \exp\left(-\frac{1}{\beta} |h_x - M| |h_y - M| (1 - s_x s_y)\right). \qquad (70)$$

The algorithm described so far is not ergodic. We therefore include in our update procedure a local Metropolis updating of the spins.

---

[11] For a discussion of the detailed balance condition see the same reference.

## 4.2 Performance of the Algorithm

We made an attempt to estimate the dynamical critical exponent of our VMR algorithm in the rough phase. To this end we measured various quantities. Local observables were

$$E = \langle (\varphi_x - \varphi_y)^2 \rangle, \quad \text{with } x, y \text{ nearest neighbours} \tag{71}$$

and

$$C = \langle \cos(2\pi\varphi_x) \rangle. \tag{72}$$

As nonlocal quantities we measured the surface thickness $\sigma$ and a set of "block spin observables". To this end, the lattice was divided into four blocks $x'$ of size $L/2 \times L/2$. Block spins $\phi_{x'}$ were then defined as the averages of $\varphi$ over these blocks. We looked at[12]

$$A_{1,2} = \frac{1}{8} \sum_{<x',y'>} \left\langle (\phi_{x'} - \phi_{y'})^2 \right\rangle, \tag{73}$$

and

$$A_{3,2} = \frac{1}{4} \sum_{x'} \langle \cos(2\pi\phi_{x'}) \rangle. \tag{74}$$

The lattice average $\Phi$ of $\varphi$ was already defined in the introduction. We also measured the quantity

$$A_{3,1} = \langle \cos(2\pi\Phi) \rangle. \tag{75}$$

The efficiency of a stochastic algorithm can be characterized by the integrated autocorrelation time

$$\tau = \frac{1}{2} \sum_{t=-\infty}^{\infty} \rho(t), \tag{76}$$

where the normalized autocorrelation function $\rho(t)$ of an observable $O$ is given by

$$\rho(t) = \frac{\langle O_i O_{i+t} \rangle - \langle O \rangle^2}{\langle O^2 \rangle - \langle O \rangle^2}. \tag{77}$$

We calculated the integrated autocorrelation times $\tau$ with a self-consistent truncation window of width $6\tau_{\max}$ for the observables specified above, where $\tau_{\max}$ was the largest autocorrelation time found in the set of quantities considered.

In the following we shall discuss the performance of the algorithm at $\beta = 0.75$ and $z = 0$ up to $z = 1.5$ in steps of $0.5$. We made runs on lattices from $L = 8$ to $L = 256$. We always made 100000 measurements of the quantities specified above, the measurements separated by an update step that consisted of the generation and flip of one H-cluster, one I-cluster, and one Metropolis sweep. The proposed change of the spins in the Metropolis step was always taken with uniform probability from the interval $(-0.5, 0.5)$.

---

[12]The notations are chosen to be consistent with the notations of ref. [3].

It turned out that the surface thickness and $A_{3,1}$ had the largest autocorrelations. The integrated autocorrelation times for these quantities are given in table 1. Another observation that might be of general interest: The autocorrelation function of $A_{3,1}$ and the autocorrelation functions of the other block observables showed a clean exponential decay already for moderately small $t$. For the other quantities we could not observe such a clean exponential decay.

For the $z$-values considered, the point $\beta = 0.75$ is definitely in the rough phase, i.e. the correlation length is infinite. The $\tau$'s are therefore expected to scale with the lattice size $L$ like $\tau \propto L^z$, where $z$ is the dynamical critical exponent (not to be confused with the fugacity).

We fitted our data to this law. When the smallest lattice sizes were discarded, we obtained fits with good $\chi^2/\text{dof}$. Our results are given in table 2. The results for the different fugacities are fairly consistent with each other. Furthermore, an exponent of order 0.3 indicates that the CSD has been drastically reduced.

Let us now give a few more details on the runs at $\beta = 0.75$ and $z = 1.0$. The clusters built with the H-algorithm had an average size (in units of $L^2$) of 0.51 for $L = 8$ to 0.33 for $L = 256$. For the clusters of the I-algorithm the average size was fairly independent of the lattice size, always close to 0.32. The acceptance rate in the Metropolis step was approximately 60 per cent throughout.

A word on the resources needed for this study: the runs with $L = 256$ needed something like 23 hours on an HP9000/735 workstation.

We did not attempt to estimate the dynamical critical exponent in the smooth phase, but we have no doubt that the performance is comparably good there.

## 5 Comparison of Perturbation Theory and Monte Carlo Results

### 5.1 Accuracy of the Perturbation Theory

We checked the accuracy of eq. (33) by a direct comparison with Monte Carlo results for the surface thickness obtained with the VMR cluster algorithm described above.

In table 1 we compare the results at $\beta = 0.75$. For $z \leq 1$ the deviations are fairly small. For $z = 1.5$ the relative error of the perturbative result is of order 1.5 per cent.

A similar statement is true for a set of $\beta$ values that we chose to be close to the roughening point $\beta_c(z)$ (compare subsection 5.2). Table 3 shows our results for the points $(z, \beta) = (0.5, 0.665)$, $(1.0, 0.700)$, and $(1.5, 0.720)$.

Table 4 shows the same comparison for three $\beta$ values in the smooth phase. As is to be expected from the fact that $\sigma_0^2$ and $\sigma_2^2$ diverge logarithmically with $L$, the approximation fails completely for large $L$: the breaking of the symmetry under global integer shifts and the finiteness of $\sigma^2$ for $L \to \infty$ are not reproduced by the perturbation theory.

## 5.2 The Critical Line

We tried to determine $\beta_c(z)$ for $z = 0.5$, 1.0 and 1.5 using the *matching* method that we introduced in [3]. This method is based on a comparison of the blocked observables of the model in question with that of the BCSOS model. The BCSOS model is exactly solved and known to undergo a KT phase transition. The basic idea of our matching procedure is to simulate the BCSOS model at its known roughening temperature. One measures various block observables, e.g., the $A_{i,l}$ introduced in subsection 4.2. One then searches for the critical coupling $\beta_c$ and a scale transformation of the SOS model under consideration such that the blocked SOS observables match with that of the critical BCSOS model.

It was demonstrated in ref. [3], that the matching with the BCSOS model can be performed by using just two block observables, e.g., $A_{1,2}$ and $A_{3,2}$. The matching conditions then fix $\beta_c$ and the scale transformation. The matching of all the other block observables is then a consequence of universality and can be demonstrated.

We tried to use the block BCSOS observables reported in [3] for a matching analysis of the SG model. We restricted ourselves to the quantities $A_{1,2}$ and $A_{3,2}$.

It turned out that the values of $A_{3,2}$ for the SG model close to the expected critical temperature are smaller than those of the BCSOS model on lattices of size up to $L = 128$ [3]. Since $A_{3,2}$ vanishes like the inverse of the logarithm of the lattice size there is no hope to simulate the BCSOS model on lattices so large that the direct matching with the SG model at the $z$ values under consideration can be performed.

From KT theory one expects that to leading order $A_{1,2}$ is a linear function of $A_{3,2}$. In figure 2 we plot for the BCSOS model the quantity $D_{1,2}$ as a function of $A_{3,2}$, where

$$D_{1,2}(L) = \frac{A_{1,2}(\infty)|_{z=0}}{A_{1,2}(L)|_{z=0}} A_{1,2}(L) . \qquad (78)$$

In [3] we demonstrated that replacing $A_{1,2}$ by $D_{1,2}$ reduces finite lattice-spacing artefacts considerably. In addition to the BCSOS results the $z = 0$ result is known. It is given by $A_{3,2} = 0$ and $A_{1,2} = 0.075425$. It seems that the linear interpolation between this $z = 0$ point and the $L = 128$ point of the BCSOS model is a reasonable approximation to the curve. On obtains $A_{1,2} = 0.075425 + 0.034(2)A_{3,2}$. $\beta_c$ is now computed from the condition that $A_{1,2}(\beta)$ and $A_{3,2}(\beta)$ are elements of the curve.

We simulated the SG model at $\beta$'s close to the expected value for $\beta_c$ on lattices up to $L = 64$. We computed the values for the two observables in the neighbourhood of this $\beta$ using the reweighting technique. Instead of $A_{1,2}$ itself we again considered $D_{1,2}$ in order to reduce finite lattice-spacing artefacts. The results for $\beta_c$ stemming from different lattice sizes are summarized in table 5. We give two error bars. The first stems from the statistical error of the SG data itself, while the second is due to the error bar of the slope of the critical $A_{1,2}(A_{3,2})$ curve. Even the result from $L = 8$ is consistent with the results from larger lattice sizes within the statistical errors.

From the exact result at $z = 0$ and the estimate $\beta_c(0.5) = 0.670(2)$ we derive a slope of the critical line of $S = 14.98(0.9)$. This is obviously inconsistent with our estimate $S = 17.945$ from the second order perturbation theory.

Beyond a three standard deviation statistical error there are two possible sources for a systematic error. First, the linear approximation of $A_{1,2}(A_{3,2})$ might be inadequate. Second, we were not able to check whether higher orders in $z$ of the surface thickness expansion can modify the slope of the critical curve. Such a modification would happen if the analogue of $A_\infty$ at order $z^n$ would diverge like $1/x^{(n-1)}$ for $x \to 0$.

## 6  Summary and Conclusion

We have presented a computation of the lines of constant physics of the SG model in the rough phase. The calculation was based on an expansion of the surface thickness to second order in the fugacity $z$. The comparison of the surface thickness data with the ones obtained with a very efficient VMR cluster algorithm revealed the reliability of the expansion in the small fugacity regime. When comparing the slope of the critical line with the result obtained from a RG matching with the BCSOS model, we found a significant deviation. We cannot presently decide whether this effect is due to the higher order terms in $z$, or whether there is another, yet undiscovered reason for the deviation.

## Acknowledgment

One of us (K.P.) would like to thank Gernot Münster for helpful discussions.

## Appendix 1: The $z = 0$ Surface Thickness

The surface thickness for $z = 0$ is given by

$$\sigma_o^2 = \frac{1}{N} \sum_x \langle (\varphi_x - \varphi_o)^2 \rangle_{\beta C} . \tag{79}$$

It is not difficult to demonstrate that

$$\sigma_o^2 = 2\beta \frac{1}{N} \sum_{p \neq 0} \frac{1}{\hat{p}^2} , \tag{80}$$

where $\hat{p}^2 = 4 - 2\cos p_1 - 2\cos p_2$, and the $p_i$, $i = 1,2$, are summed over the values $\{0, ..., L-1\} \cdot (2\pi/L)$.

We want to prove the following

STATEMENT: For $L \to \infty$, the surface thickness $\sigma_o^2$ goes like

$$\sigma_o^2 \to \frac{\beta}{\pi} \ln L + c . \tag{81}$$

For the proof of the statement we need the following definitions (for formal reasons let us assume that $L$ is even):

$$\Lambda_L = \left\{ p = (p_1, p_2) : p_i = n_i \frac{2\pi}{L}, \ -(\frac{L}{2} - 1) \leq n_i \leq \frac{L}{2} \right\} \tag{82}$$

and

$$\Lambda_L^\infty = \left\{ p = (p_1, p_2) : p_i = n_i \frac{2\pi}{L}, \ -\infty \leq n_i \leq \infty \right\} . \tag{83}$$

Now we define the quantities

$$Q(L) = \frac{1}{L^2} \sum_{p \in \Lambda_L \setminus \{0\}} \frac{1}{\hat{p}^2} \tag{84}$$

and

$$R(L, a) = \frac{1}{L^2} \sum_{p \in \Lambda_L^\infty \setminus \{0\}} \frac{e^{-a^2 p^2}}{p^2} . \tag{85}$$

Note that $a$ provides an ultraviolet cut-off in the second sum. To prove the statement we need the following

LEMMA: For $a > 0$, we have

$$\lim_{L \to \infty} \{Q(L) - R(L, a)\} < \infty . \tag{86}$$

PROOF OF THE LEMMA: We write the difference of $Q$ and $R$ as a sum of two contributions,

$$Q(L) - R(L, a) = \underbrace{\frac{1}{L^2} \sum_{p \in \Lambda_L \setminus \{0\}} \left\{ \frac{1}{\hat{p}^2} - \frac{1}{p^2} \right\}}_{T_1} - \underbrace{\frac{1}{L^2} \sum_{p \in \Lambda_L^\infty \setminus \Lambda_L} \frac{e^{-a^2 p^2}}{p^2}}_{T_2} . \tag{87}$$

In the limit $L \to \infty$, the first term becomes

$$T_1 \to \int_{-\pi}^{\pi} \frac{d^2 p}{(2\pi)^2} \left\{ \frac{1}{\hat{p}^2} - \frac{1}{p^2} \right\} < \infty , \tag{88}$$

because

$$\lim_{p \to 0} \left\{ \frac{1}{\hat{p}^2} - \frac{1}{p^2} \right\} = \text{ finite} . \tag{89}$$

Let us define

$$G := \{p : |p_i| \geq \pi\} . \tag{90}$$

In the limit of large $L$ the second term becomes

$$T_2 \to \int_G \frac{d^2 p}{(2\pi)^2} \frac{e^{-a^2 p^2}}{p^2} < \frac{1}{2\pi} \int_\pi^\infty dq \frac{e^{-a^2 q^2}}{q} < \infty . \tag{91}$$

It follows from the Lemma that the parts of $Q(L)$ and $R(L,a)$ that diverge when $L \to \infty$ are identical (since their difference stays finite in this limit). Employing the standard "Feynman" trick, we get

$$\begin{aligned} R(L,a) &= \frac{1}{L^2} \sum_{p \in \Lambda_L^\infty \setminus \{0\}} \frac{e^{-a^2 p^2}}{p^2} \\ &= \frac{1}{L^2} \int_0^\infty dt \, \{ \underbrace{\sum_{p \in \Lambda_L^\infty} e^{-(t+a^2)p^2}}_{A(\frac{4\pi}{L^2}(t+a^2))^2} - 1 \}. \end{aligned} \qquad (92)$$

Here, $A$ denotes the Jacobi-$\vartheta$-function,

$$A(s) := \sum_{n \in \mathbf{Z}} e^{-\pi n^2 s}. \qquad (93)$$

After two simple substitutions we find

$$R(L,a) = \frac{1}{4\pi} \int_{\frac{4\pi a^2}{L^2}}^\infty dt \, \{A(t)^2 - 1\} \qquad (94)$$

For $t \to 0$ we have

$$A(t) \to \sqrt{\frac{1}{t}} \{1 + e^{-\frac{\pi}{t}} + ...\}. \qquad (95)$$

It is therefore instructive to split the integral representation of $R$ as follows:

$$\begin{aligned} R(L,a) &= \frac{1}{4\pi} \int_{\frac{4\pi a^2}{L^2}}^1 dt \, \{A(t)^2 - \frac{1}{t}\} \\ &+ \frac{1}{4\pi} \int_1^\infty dt \, \{A(t)^2 - 1\} \\ &+ \frac{1}{4\pi} \int_{\frac{4\pi a^2}{L^2}}^1 dt \, \{\frac{1}{t} - 1\}. \end{aligned} \qquad (96)$$

The first two terms stay finite when $L \to \infty$, the third integral diverges logarithmically:

$$R(L,a) \to \text{finite} + \frac{1}{2\pi} \ln L. \qquad (97)$$

This is what we wanted to show.

## Appendix 2: Gaussian Measures

Expectation values in the Sine Gordon model are defined by

$$\langle \mathcal{O}(\varphi) \rangle = \lim_{M \to 0} \frac{\int \prod_x d\varphi_x \exp[-H_M(\varphi)] \, \mathcal{O}(\varphi)}{\int \prod_x d\varphi_x \exp[-H_M(\varphi)]}. \qquad (98)$$

Here, we have introduced a symmetry breaking "mass" term that serves temporarily as a regulator of the zero mode:

$$H_M(\varphi) = \frac{1}{2\beta} \sum_{<x,y>} (\varphi_x - \varphi_y)^2 + \frac{M^2}{2} \sum_x \varphi_x^2 - V(\varphi) \,. \tag{99}$$

We define the scalar product

$$(\phi, \psi) = \sum_x \phi_x \psi_x \,, \tag{100}$$

and a lattice Laplacian $\Delta$,

$$(\Delta \varphi)_x = \sum_{y.nn.x} (\varphi_y - \varphi_x) \,, \tag{101}$$

where the sum runs over the sites $y$ that are nearest neighbours of $x$. Eq. (99) can now be rewritten as

$$H_M(\varphi) = \frac{1}{2} \left( \varphi, \left[ -\frac{\Delta}{\beta} + M^2 \right] \varphi \right) - V(\varphi) \,. \tag{102}$$

Now we write

$$\langle \mathcal{O}(\varphi) \rangle = \lim_{M \to 0} \frac{\int \prod_x d\varphi_x \exp\left[ -\frac{1}{2}(\varphi, v_M^{-1} \varphi) + V(\varphi) \right] \mathcal{O}(\varphi) \Big/ \int \prod_x d\varphi_x \exp\left[ -\frac{1}{2}(\varphi, v_M^{-1} \varphi) \right]}{\int \prod_x d\varphi_x \exp\left[ -\frac{1}{2}(\varphi, v_M^{-1} \varphi) + V(\varphi) \right] \Big/ \int \prod_x d\varphi_x \exp\left[ -\frac{1}{2}(\varphi, v_M^{-1} \varphi) \right]} \,. \tag{103}$$

Here, we have introduced the abbreviation

$$v_M^{-1} = -\frac{\Delta}{\beta} + M^2 \,. \tag{104}$$

The matrix $v_M$ has an inverse $v_M$ that can be explicitly calculated by Fourier transformation:

$$v_{M,xy} = \frac{1}{N} \sum_p \frac{e^{ip(x-y)}}{\hat{p}^2/\beta + M^2} \,. \tag{105}$$

Here, the $p_i$, $i = 1, 2$, run over the values $\{0, ..., L-1\} \cdot (2\pi/L)$, and

$$\hat{p}^2 = 4 - 2\cos p_1 - 2\cos p_2 \,. \tag{106}$$

The Gaussian measure with covariance $v_M$ is defined through

$$\int d\mu_{v_M}(\varphi) \, \mathcal{O}(\varphi) = \frac{\int \prod_x d\varphi_x \exp\left[ -\frac{1}{2}(\varphi, v_M^{-1} \varphi) \right] \mathcal{O}(\varphi)}{\int \prod_x d\varphi_x \exp\left[ -\frac{1}{2}(\varphi, v_M^{-1} \varphi) \right]} \,. \tag{107}$$

With this definition we can write

$$\langle \mathcal{O}(\varphi) \rangle = \lim_{M \to 0} \frac{\int d\mu_{v_M}(\varphi) \exp\left[ V(\varphi) \right] \mathcal{O}(\varphi)}{\int d\mu_{v_M}(\varphi) \exp\left[ V(\varphi) \right]} \,. \tag{108}$$

The generating functional of this measure can be explicitly calculated:

$$\int d\mu_{v_M}(\varphi) \exp\left[i(k,\varphi)\right] = \exp\left[-\tfrac{1}{2}(k, v_M k)\right]. \qquad (109)$$

All correlation functions can be derived from this relation. Let us now see what happens in the limit $M \to 0$. We split

$$v_{M,xy} = \frac{1}{N} \sum_p \frac{e^{ip(x-y)} - 1}{\hat{p}^2/\beta + M^2} + \frac{1}{N} \sum_p \frac{1}{\hat{p}^2/\beta + M^2}. \qquad (110)$$

The first term stays finite when $M \to 0$, the second (to be called $K(M)$) diverges to $+\infty$. Therefore

$$\begin{aligned}
\int d\mu_{\beta C}(\varphi) \exp\left[i(k,\varphi)\right] &\equiv \lim_{M \to 0} \int d\mu_{v_M}(\varphi) \exp\left[i(k,\varphi)\right] \\
&= \exp\left[-\tfrac{1}{2}\beta(k, Ck)\right] \lim_{M \to 0} \exp\left[-\tfrac{1}{2} K(M) \left(\sum_x k_x\right)^2\right] \\
&= \begin{cases} \exp\left[-\tfrac{1}{2}\beta(k, Ck)\right] & , \text{if } \sum_x k_x = 0 \\ 0 & , \text{else}. \end{cases}
\end{aligned} \qquad (111)$$

The matrix elements of $C$ are given by

$$C_{xy} = \frac{1}{N} \sum_{p \neq 0} \frac{e^{ip(x-y)} - 1}{\hat{p}^2}. \qquad (112)$$

# References

[1] To give only a few references:
J.M. Kosterlitz and D.J. Thouless, J. Phys. C6, 1181 (1973);

J.M. Kosterlitz, J. Phys. C7, 1046 (1974);

S.T. Chui and J.D. Weeks, Phys. Rev. B14, 4978 (1976);

J.V. José, L.P. Kadanoff, S. Kirkpatrick, and D.R. Nelson,
Phys. Rev. B16, 1217 (1977);

T. Ohta and K. Kawasaki, Prog. Theor. Phys. 60, 365 (1978);

D.J. Amit, Y.Y. Goldschmidt, and G. Grinstein, J. Phys. A13, 585 (1980).

[2] D.B. Abraham, 'Surface Structures and Phase Transitions – Exact Results', in: Phase Transitions and Critical Phenomena, Vol. 10, C. Domb and J.L. Lebowitz, eds., Academic 1986;

H. van Beijeren and I. Nolden, 'The Roughening Transition', in: Topics in Current Physics, Vol. 43, W. Schommers and P. van Blankenhagen, eds., Springer 1987.

[3] M. Hasenbusch, M. Marcu, and K. Pinn, Physica A 208, 124 (1994).

[4] S. Samuel, Phys. Rev. D 18, 1916 (1978).

[5] H.G. Evertz, M. Hasenbusch, M. Marcu, K. Pinn, and S. Solomon,
Phys. Lett. B254, 185 (1991);

H.G. Evertz, M. Hasenbusch, G. Lana, M. Marcu, and K. Pinn,
Phys. Rev. B 46, 10472 (1992).

[6] R.H. Swendsen and J.S. Wang, Phys. Rev. Lett. 58, 86 (1987).

[7] U. Wolff, Phys. Rev. Lett. 62, 361 (1989).

[8] C. Itzykson, J.M. Drouffe, 'Statistical Field Theory', Cambridge University Press 1989.

[9] G.H. Hardy, Mess. Math. 49, 85 (1919).

[10] W. Magnus, F. Oberhettinger, F.G. Tricomi, 'Higher Transcendental Functions', Vol. 1, A. Erdélyi, ed., McGraw-Hill, New York, 1953.

[11] G. Münster, private communication.

| $z$ | $L$ | $A_{3,1}$ | $\tau_{A_{3,1}}$ | $\tau_{\sigma^2}$ | $\sigma^2_{\text{MC}}$ | $\sigma^2_{\text{PT}}$ |
|-----|-----|-----------|------------------|-------------------|------------------------|------------------------|
| 0.0 | 8   | 0.0087(57)  | 3.27(07)  | 1.37(4)  | 0.56751(75) | 0.568942 |
| 0.0 | 16  | -0.0085(70) | 4.78(12)  | 1.75(5)  | 0.73483(81) | 0.734887 |
| 0.0 | 32  | 0.0000(81)  | 6.56(18)  | 2.41(11) | 0.89919(92) | 0.900489 |
| 0.0 | 64  | -0.0005(87) | 7.53(24)  | 2.72(13) | 1.06513(98) | 1.065998 |
| 0.0 | 128 | 0.007(10)   | 9.71(41)  | 3.51(18) | 1.2324(11)  | 1.231483 |
| 0.0 | 256 | 0.009(11)   | 11.58(39) | 3.78(21) | 1.3972(12)  | 1.396961 |
| 0.5 | 8   | 0.0571(58)  | 3.38(08)  | 1.45(5)  | 0.56602(78) | 0.566367 |
| 0.5 | 16  | 0.0539(70)  | 4.83(13)  | 1.73(5)  | 0.73104(80) | 0.731812 |
| 0.5 | 32  | 0.0258(79)  | 6.19(19)  | 2.31(10) | 0.89795(92) | 0.896745 |
| 0.5 | 64  | 0.0307(86)  | 7.35(22)  | 2.63(11) | 1.05992(95) | 1.061484 |
| 0.5 | 128 | 0.0162(97)  | 9.25(38)  | 3.10(16) | 1.2264(10)  | 1.226138 |
| 0.5 | 256 | 0.005(10)   | 10.67(41) | 3.89(17) | 1.3925(12)  | 1.390748 |
| 1.0 | 8   | 0.1021(57)  | 3.30(07)  | 1.61(4)  | 0.56079(85) | 0.558644 |
| 1.0 | 16  | 0.0849(69)  | 4.69(10)  | 1.96(6)  | 0.72417(87) | 0.722587 |
| 1.0 | 32  | 0.0694(78)  | 6.05(22)  | 2.31(7)  | 0.88976(92) | 0.896745 |
| 1.0 | 64  | 0.0518(89)  | 7.91(28)  | 2.69(12) | 1.05077(96) | 1.061484 |
| 1.0 | 128 | 0.0370(91)  | 8.31(33)  | 3.42(16) | 1.2148(11)  | 1.226138 |
| 1.0 | 256 | 0.023(11)   | 11.29(39) | 3.98(19) | 1.3776(12)  | 1.372110 |
| 1.5 | 8   | 0.1466(57)  | 3.34(08)  | 1.62(4)  | 0.55412(86) | 0.545772 |
| 1.5 | 16  | 0.1114(69)  | 4.79(13)  | 1.92(7)  | 0.71853(86) | 0.707212 |
| 1.5 | 32  | 0.0984(77)  | 5.91(17)  | 2.36(8)  | 0.87879(92) | 0.866793 |
| 1.5 | 64  | 0.0779(90)  | 8.02(24)  | 2.73(13) | 1.04003(97) | 1.025370 |
| 1.5 | 128 | 0.0786(97)  | 9.28(27)  | 3.69(20) | 1.2015(11)  | 1.183376 |
| 1.5 | 256 | 0.0634(99)  | 10.00(41) | 3.82(20) | 1.3594(11)  | 1.341046 |

Table 1: Monte Carlo estimates for the quantities $A_{3,1}$ and $\sigma^2$ for different lattice sizes $L$ and fugacities $z$. $\beta$ is 0.75 always. The Monte Carlo results for the surface thickness squared can be compared with leading order perturbation theory (PT) given in the last column. The $\tau$'s give the integrated autocorrelation times.

| $z$ | $z_{A_{3,1}}$ | $z_{\sigma^2}$ |
|-----|---------------|----------------|
| 0.0 | 0.282(18)     | 0.237(28)      |
| 0.5 | 0.290(15)     | 0.285(17)      |
| 1.0 | 0.310(13)     | 0.257(18)      |
| 1.5 | 0.294(15)     | 0.264(20)      |

Table 2: Estimates for the dynamical critical exponents as obtained from the integrated autocorrelation times of the quantities $A_{3,1}$ and $\sigma^2$.

| $z$ | $\beta$ | $L$ | $\sigma^2_{\mathrm{MC}}$ | $\sigma^2_{\mathrm{PT}}$ |
|-----|---------|-----|-------------------------|--------------------------|
| 0.5 | 0.665 | 8  | 0.49914(41) | 0.498831 |
| 0.5 | 0.665 | 16 | 0.64426(39) | 0.644232 |
| 0.5 | 0.665 | 32 | 0.78950(40) | 0.788863 |
| 0.5 | 0.665 | 64 | 0.93364(39) | 0.933018 |
| 1.0 | 0.700 | 8  | 0.51899(45) | 0.514735 |
| 1.0 | 0.700 | 16 | 0.66990(42) | 0.665522 |
| 1.0 | 0.700 | 32 | 0.81866(42) | 0.814886 |
| 1.0 | 0.700 | 64 | 0.96894(42) | 0.963367 |
| 1.5 | 0.720 | 8  | 0.52654(48) | 0.515716 |
| 1.5 | 0.720 | 16 | 0.68148(44) | 0.668146 |
| 1.5 | 0.720 | 32 | 0.83422(43) | 0.818267 |
| 1.5 | 0.720 | 64 | 0.98679(43) | 0.966976 |

Table 3: Monte Carlo estimates for the surface thickness close to $\beta_c(z)$, and comparison with perturbation theory.

| $z$ | $\beta$ | $L$ | $\sigma^2_{\mathrm{MC}}$ | $\sigma^2_{\mathrm{PT}}$ |
|---|---|---|---|---|
| 0.5 | 0.60 | 8 | 0.44624(69) | 0.444773 |
| 0.5 | 0.60 | 16 | 0.57277(74) | 0.572703 |
| 0.5 | 0.60 | 32 | 0.69975(84) | 0.698515 |
| 0.5 | 0.60 | 64 | 0.82352(93) | 0.822110 |
| 1.0 | 0.60 | 8 | 0.42250(83) | 0.413632 |
| 1.0 | 0.60 | 16 | 0.53894(88) | 0.527084 |
| 1.0 | 0.60 | 32 | 0.64558(98) | 0.632884 |
| 1.0 | 0.60 | 64 | 0.7416(12) | 0.730045 |
| 1.5 | 0.60 | 8 | 0.39822(90) | 0.361730 |
| 1.5 | 0.60 | 16 | 0.50207(96) | 0.451052 |
| 1.5 | 0.60 | 32 | 0.5862(12) | 0.523500 |
| 1.5 | 0.60 | 64 | 0.6481(11) | 0.576603 |
| 0.5 | 0.50 | 8 | 0.35450(67) | 0.352331 |
| 0.5 | 0.50 | 16 | 0.44429(81) | 0.438912 |
| 0.5 | 0.50 | 32 | 0.5167(10) | 0.504380 |
| 0.5 | 0.50 | 64 | 0.5639(11) | 0.531686 |
| 1.0 | 0.50 | 8 | 0.30370(83) | 0.271438 |
| 1.0 | 0.50 | 16 | 0.35516(83) | 0.285876 |
| 1.0 | 0.50 | 32 | 0.38203(71) | 0.216543 |
| 1.0 | 0.50 | 64 | 0.38900(42) | -0.005254 |
| 1.5 | 0.50 | 8 | 0.25865(84) | 0.136617 |
| 1.5 | 0.50 | 16 | 0.28566(66) | 0.030816 |
| 1.5 | 0.50 | 32 | 0.29718(43) | -0.263186 |
| 1.5 | 0.50 | 64 | 0.29994(22) | -0.900152 |

Table 4: Monte Carlo estimates for the surface thickness in the smooth phase, and comparison with perturbation theory.

| $z$ | $L = 8$ | $L = 16$ | $L = 32$ | $L = 64$ | $\beta_c(z)$ |
|---|---|---|---|---|---|
| 0.0 | | | | | 0.63662 |
| 0.5 | 0.6685(16)(12) | 0.6709(13)(9) | 0.6719(11)(8) | 0.6683(18)(9) | 0.670(2) |
| 1.0 | 0.6989(13)(15) | 0.6955(11)(13) | 0.6984(9)(10) | 0.6957(12)(8) | 0.697(2) |
| 1.5 | 0.7153(13)(16) | 0.7137(10)(13) | 0.7112(10)(10) | 0.7105(12)(10) | 0.711(2) |
| $\infty$ | | | | | 0.7524(7) |

Table 5: Estimates for the critical couplings $\beta_c(z)$ from matching of $A_{1,2}$ and $A_{3,2}$. $\beta_c$ for $z = \infty$ is the roughening coupling for the Discrete Gaussian model [3].

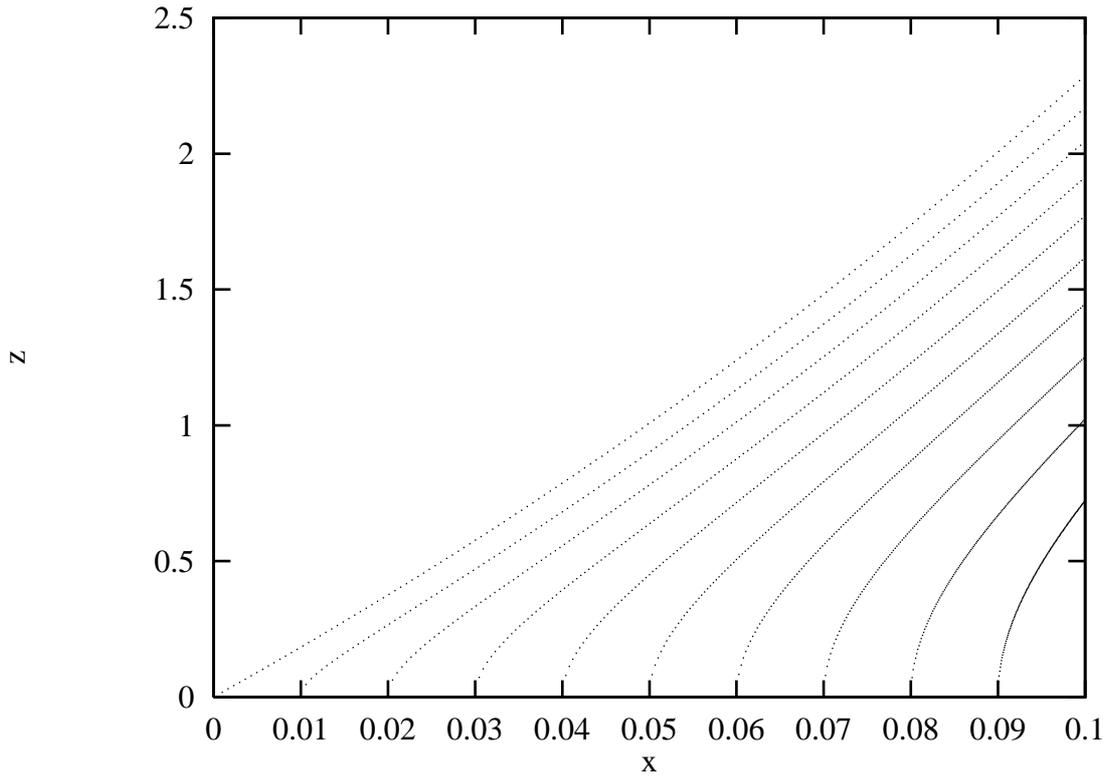

Figure 1: Lines of constant physics in the rough phase of the 2-dimensional lattice Sine Gordon model for $x_{\text{eff}} = 0.0 \ldots 0.09$.

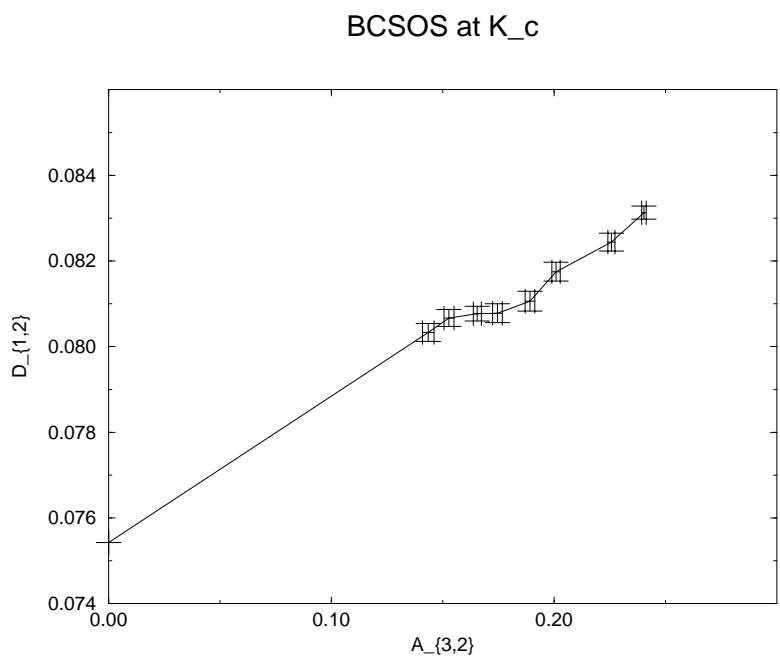

Figure 2: The quantity $D_{1,2}$ as a function of $A_{3,2}$ for the BCSOS model at criticality. The data are taken from ref. [3].